\documentclass[aps,pra,twocolumn,superscriptaddress,longbibliography]{revtex4-2} 

\usepackage{graphicx}
\usepackage{dcolumn}
\usepackage{bm}
\usepackage{times} 
\usepackage{amsmath,amssymb}
\usepackage{float}
\usepackage{color}
\usepackage{multirow}
\usepackage[normalem]{ulem}

\begin{document}
\title{The infinitesimal environmental dust as a photonic bath at infinity}

\author{Gaomin Tang}
\affiliation{Graduate School of China Academy of Engineering Physics, Beijing 100193,
China}
\author{Yuhua Ren}
\affiliation{Department of Physics, National University of Singapore, Singapore 117551,
Republic of Singapore}
\author{Jian-Sheng Wang}
\affiliation{Department of Physics, National University of Singapore, Singapore 117551,
Republic of Singapore}

\bigskip

\begin{abstract}
  In far-field thermal radiation, electromagnetic waves emitted by an object propagate to
  infinity, requiring the far region to be modeled as an effective thermal bath.
  This bath was proposed as infinitesimal environmental ``dust", but explicit calculations
  with such distributed dust involve integrals over infinite space that are difficult to
  evaluate.
  In this work, we map this environmental dust to a photonic bath at infinity within the
  nonequilibrium photonic Green's function formalism. By explicitly evaluating the spatial
  integral over the dust, we show that its contribution reduces to a simple local
  self-energy, for which we derive analytical expressions for both three-dimensional
  objects and planar systems. We further demonstrate that the bath behaves as a black body
  and clarify its role in far-field thermal radiation. 
  An alternative derivation based on the surface Green's function framework is also
  provided, demonstrating the theoretical consistency of the results without invoking the
  dust model. The photonic bath at infinity provides a convenient framework for both
  analytical and numerical calculations in far-field thermal radiation.
\end{abstract}

\maketitle

\section{Introduction}
In far-field thermal radiation, electromagnetic waves emitted by an object ultimately
propagate to infinity, where they are absorbed or dissipated. A physically consistent
description of this process requires modeling the far region as an effective thermal bath.
Following the interpretation of Ref.~\cite{Eckhardt84}, the environment can be modeled as
``dust" medium with a homogeneous infinitesimal dielectric response occupying the infinite
space outside the object. This picture is conceptually attractive, yet performing explicit
analytical or numerical calculations with such a distributed dust is far from trivial, as
it involves integrals over the entire infinite volume.

The environmental contribution has been evaluated for a cylindrical
object~\cite{Kruger11_PRL, Kruger12_PRE, Kruger12} and for a slab of finite thickness with
in-plane translational invariance~\cite{Henne96, Henne08, Henne09}. These studies showed
that, although the dust is infinitesimally weak, its integration over infinite space
yields finite and physically meaningful results. Recently, it has been demonstrated that
the infinitesimal environmental dust can be identified as a photonic bath at infinity
within the nonequilibrium Green's function formalism~\cite{JSW22, JSW23, GT25, Zhu26}.
This identification replaces the otherwise cumbersome integral over infinite space with a
simple local self-energy, greatly simplifying both analytical and numerical treatments.

The introduction of a photonic bath at infinity offers conceptual and technical advantages
when using nonequilibrium Green's functions to describe radiative energy and momentum
transfer, as well as the Casimir force~\cite{Henne08, Henne09, Jiebin17, JSW17-EPL, JSW20,
JSW22, Basko22, JSW23, JSW24, JSW25, GT25, Belzig25, Zhu26, JSW26, JSW26-2, GT26-2}.
One of the motivations is to resolve a formalism inconsistency regarding zero-point
fluctuations.
While employing normal ordering for the Poynting vector in local thermal equilibrium
yields a net heat current proportional solely to the difference in Bose-Einstein
distribution functions~\cite{Jiebin17, JSW17-EPL, JSW23}, applying the same procedure to
the Maxwell stress tensor causes the Casimir effect to vanish.  This occurs because the
Casimir force originates from zero-point fluctuations, which are explicitly removed by
normal ordering~\cite{JSW23}.
This challenge is addressed by using the symmetrized Keldysh components of the photonic
Green's function and the self-energy, which naturally incorporate the Bose-Einstein
distribution and zero-point contributions. In this framework, the photonic bath at
infinity accounts for the absorption and emission of vacuum fluctuations, leading to a
natural cancellation of the zero-point terms in far-field thermal radiation.
Crucially, this formulation preserves the zero-point fluctuations necessary for the
Casimir force, which cannot be reduced to a simple Landauer-B\"{u}ttiker-like expression
for forces~\cite{JSW23, JSW25}. Furthermore, the bath at infinity is essential to preserve
conservation laws for far-field heat, momentum, and angular-momentum
radiation~\cite{Jiebin17, JSW20, JSW22, JSW23, GT25, Zhu26}.

Beyond static systems, the photonic bath at infinity plays a critical role in describing
far-field thermal radiation in time-modulated (Floquet) systems~\cite{Lozano23, GT25,
Zhu26, JSW26} as well as in nonstationary transient regimes~\cite{Kiryl25}. For
transient systems, the bath accounts for continuous real-time energy dissipation into the
infinite vacuum. For a Floquet-driven system, standard normal ordering predicts zero
energy transfer rate at zero temperature, as the equilibrium distribution $N(\omega) =
-\theta(-\omega)$ suggests that positive-frequency modes have no thermal population.
However, Floquet driving provides external work that can shift fluctuations from the
negative-frequency domain into the positive domain, giving rise to physical emission even
at zero temperature~\cite{YuFan25, Zhu26}.
This emission is correctly captured by the photonic bath at infinity, which acts as the
necessary sink for these shifted fluctuations.
In this context, the bath ensures that the formalism accounts for the conversion of
modulation work into far-field radiation, maintaining global energy conservation between
the objects, the driving source, and the environment~\cite{Zhu26}.

In this work, we systematically develop the theory of the photonic bath at infinity within
the nonequilibrium Green's function formalism. By analytically evaluating the
infinite-volume integrals of the continuum dust model and providing an independent surface
Green's function derivation, we show that the infinite electromagnetic vacuum is
equivalent to a universal boundary self-energy. This exact formulation yields closed-form
expressions for both three-dimensional and planar geometries, establishing the photonic
bath at infinity as the electromagnetic analog of the lead self-energy in quantum
transport. We further demonstrate that the bath at infinity acts as a black body, and
clarify its role in far-field thermal radiation.

The paper is structured as follows. We begin with the introduction of the photonic Keldysh
Green's function and establish the theoretical framework in Sec.~\ref{SecII}.
Section~\ref{SecIII} develops the three-dimensional photonic bath at infinity, with the
corresponding self-energy given in Eq.~\eqref{Pi_infty_3D}.
Section~\ref{SecIV} treats the two-dimensional case with in-plane translational
invariance, where the self-energy is given in Eq.~\eqref{Pi_infty_2D}.
In addition, we provide a complementary derivation in Sec.~\ref{SecV} using the surface
Green's function approach, demonstrating the framework's consistency without relying on
the dust model. Our work is summarized in Sec.~\ref{SecVI}.

\section{Photonic Green's functions} \label{SecII}
\begin{figure}
\centering
\includegraphics[width=\columnwidth]{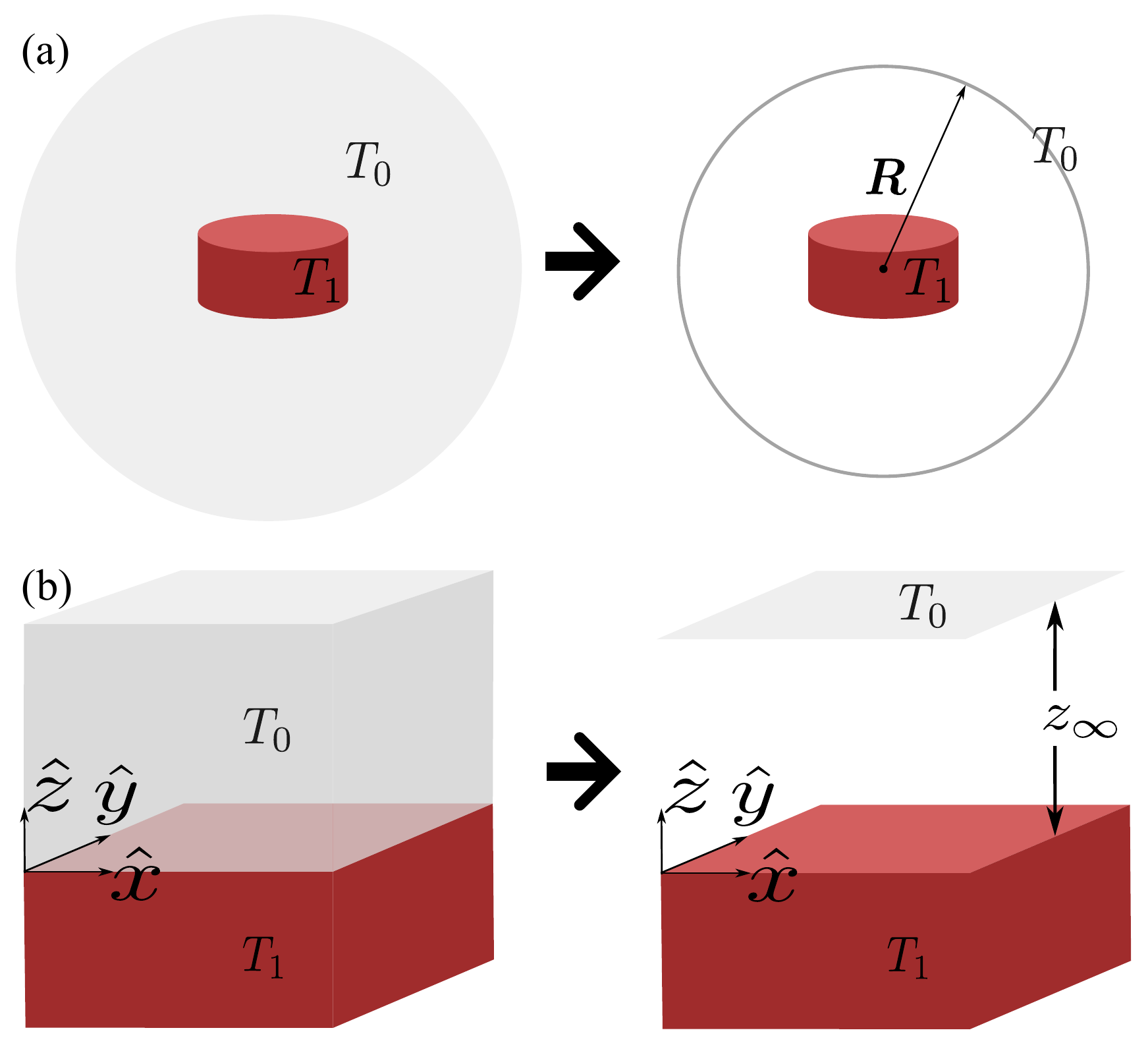} \\
\caption{Schematic of thermal radiation from (a) a three-dimensional object of finite size
  and (b) a semi-infinite object filling the region of $z<0$ to the surrounding
  environment. Both objects are maintained at temperature $T_1$, while the surrounding
  environment is at temperature $T_0$.
  The left panels depict that the environment extending to infinity is modeled as
  infinitesimal ``dust" with homogeneous dielectric properties modeled by self-energy
  $\Pi^r_0$ (gray shading). The right panels show the equivalent theoretical
  representation in which the effect of the distributed dust is modeled as a photonic bath
  at infinity with effective self-energy $\Pi^r_\infty$. }
\label{fig1}
\end{figure}

We consider thermal radiation emitted by an object at temperature $T_1$ into its
surrounding environment at temperature $T_0$, as illustrated in the left panel of
Fig.~\ref{fig1}.
The object in this work can be either a three-dimensional body of finite size
[Fig.~\ref{fig1}(a)] or a semi-infinite system with translational invariance in the
$x$-$y$ plane [Fig.~\ref{fig1}(b)].
Although we focus on a single object for clarity, the theoretical framework developed here
can be readily extended to systems involving multiple objects.
The central goal of this work is to model the environment, which extends to
infinity and possesses infinitesimal absorption, as a photonic bath at infinity.

The interaction between the material system and the electromagnetic field is described
within the minimal coupling by
\begin{equation} \label{H_int}
  H_{\rm int} = - \int d^3\bm{r} \bm{j} \cdot \bm{A} ,
\end{equation}
where $\bm{A}$ is the vector potential operator and $\bm{j}$ is the total current density
operator in the object. 
To eliminate gauge redundancy, we adopt the temporal gauge by setting the scalar potential
to zero~\cite{Lifshitz_book, QFT-Fradkin, temp2021}. In this gauge, the electric and
magnetic fields are given by 
\begin{equation} \label{EABA}
  \bm{E}=-\partial_t \bm{A}, \qquad \bm{B}=\nabla \times \bm{A}, 
\end{equation}
respectively, which simplifies the expressions for energy density and energy flux.
Within linear-response theory, the total current is decomposed as $\bm{j} =
\bm{j}_{\rm fluc} + \bm{j}_{\rm ind}$, where $\bm{j}_{\rm fluc}$ is the stochastic current
driven by background thermal fluctuations, and $\bm{j}_{\rm ind}$ represents the
field-induced current incorporating many-body dielectric screening.

The photonic Green's function on the Keldysh contour is defined via the vector potential
$\bm{A}$ in the Heisenberg picture as~\cite{Lifshitz_book, JSW23} 
\begin{equation} \label{D_Keldysh}
  D_{\mu\nu}(\bm{r},\tau; \bm{r}',\tau')
  = \frac{1}{i\hbar} \left< {\cal T} A_{\mu}(\bm{r},\tau) A_{\nu}(\bm{r}',\tau') \right> ,
\end{equation}
where ${\cal T}$ is the contour time-ordering operator, $\hbar$ is the reduced Planck
constant, and $\mu,\nu = x, y, z$ denote Cartesian components.
In this work, time variables denoted by Greek letters reside on the Keldysh contour, while
those by Latin letters are real times. The real-time lesser and greater components, which
describe the statistical correlations, are defined respectively as
\begin{align}
  D^<_{\mu\nu}(\bm{r},t; \bm{r}',t') 
  &= \frac{1}{i\hbar} \left< A_\nu(\bm{r}', t')A_\mu(\bm{r},t)\right> , \label{D<} \\
  D^>_{\mu\nu}(\bm{r},t; \bm{r}',t')
  &= \frac{1}{i\hbar} \left< A_\mu(\bm{r}, t)A_\nu(\bm{r}',t')\right> .
\end{align}
The Keldysh component is defined as the sum of them~\cite{Keldysh64}:
\begin{equation} 
  D^K(\bm{r},t; \bm{r}',t') = D^<(\bm{r},t; \bm{r}',t') + D^>(\bm{r},t; \bm{r}',t') .
\end{equation}
The retarded and advanced components are defined by
\begin{equation}
  D^r(\bm{r},t;\bm{r}',t') = \theta(t-t') \big[ D^>(\bm{r},t; \bm{r}',t') -
  D^<(\bm{r},t; \bm{r}',t')  \big] , 
\end{equation}
and 
\begin{equation}
  D_{\mu\nu}^a(\bm{r},t;\bm{r}',t') = D_{\nu\mu}^r(\bm{r}',t';\bm{r},t) .
\end{equation}
The symmetries of the photonic Green's function are given in Appendix~\ref{Appendix_A}.
Within the linear-response regime, the expectation value of the vector potential induced
by fluctuational current is given by the convolution of the retarded Green's function with
the source current~\cite{Lifshitz_book}
\begin{equation}
  \langle \bm{A}(\bm{r}, t) \rangle = - \int d^3\bm{r}' \int dt' D^r(\bm{r}, t;
  \bm{r}', t') \bm{j}_{\rm fluc}(\bm{r}', t'),
\end{equation}
or in the frequency domain for time-translationally invariant system,
\begin{equation}
  \langle \bm{A}(\bm{r}, \omega) \rangle = - \int d^3\bm{r}' D^r(\bm{r}, \bm{r}', \omega)
  \bm{j}_{\rm fluc}(\bm{r}', \omega).
\end{equation}
With these definitions, we now proceed to construct the explicit form of the photonic
Green's function, first for the free space and then in the presence of an object.

\subsection{Green's function in the free space}
We first examine the photonic Green's function in free vacuum, in the absence of any
objects. In the temporal gauge, the vector potential $\bm{A}$ satisfies the inhomogeneous
vector Helmholtz equation derived from Maxwell's equations with
\begin{equation} \label{A_EOM}
  v^{-1} \bm{A} \equiv -\left(\epsilon_0 \partial_t^2 + \mu_0^{-1} \nabla\times\nabla
  \times \right) \bm{A} = -\bm{j} ,
\end{equation}
where $\epsilon_0$ is the vacuum permittivity and $\mu_0$ the vacuum permeability.
For a system with translational invariance in space and time, we perform Fourier transform
with $\partial_t \to -i\omega$ and $\nabla \to i\bm{Q}$. 
Here, $\bm{Q} = (q_x, q_y, q_z)$ denotes the three-dimensional wavevector to
distinguish it from the two-dimensional in-plane wavevector $\bm{q} = (q_x, q_y)$
introduced later in Sec.~\ref{SecIV}.
The double curl operator with the identity 
$\nabla\times (\nabla \times \bm{A}) = \nabla (\nabla\cdot \bm{A}) - \nabla^2 \bm{A}$
becomes $(Q^2 \mathbb{I}- \bm{Q} \bm{Q})\bm{A}$ with $\mathbb{I}$ the unit
dyadic (identity matrix) and $\bm{Q} \bm{Q}$ denoting the dyadic product.
The Fourier transform of Eq.~\eqref{A_EOM} then yields $\bm{A} = - v \bm{j}$,
where the free photon Green's function $v$ is given by
\begin{equation} \label{v_qw}
  v(\bm{Q},\omega) = \frac{\mathbb{I} - \bm{Q} \bm{Q}/k_0^2}{\epsilon_0 \big(
  \omega^2 - c^2 Q^2 \big)} , 
\end{equation}
with $k_0=\omega /c$ and $c$ the speed of light in vacuum. This Green's function describes
the wave propagation in an ideal and non-dissipative vacuum. Its poles lie on the real
frequency axis at $\omega = \pm cQ$, reflecting the absence of any absorption.

To obtain the causal (retarded) Green's function required for linear-response theory, we
shift the pole to the lower half complex $\omega$-plane by adding $i\eta$ with $\eta$ an
infinitesimal positive number. The retarded bare Green's function is thus
\begin{equation} \label{D0_eta}
  D_0^r(\bm{Q},\omega) = \frac{\mathbb{I}-\bm{Q} \bm{Q}/\tilde{k}_0^2}{\epsilon_0
  \big[ (\omega + i\eta)^2- c^2 Q^2\big]}, 
\end{equation}
with $\tilde{k}_0=(\omega + i\eta) /c$. Physically, this corresponds to modeling the
environment as a medium containing uniform and absorbing ``infinitesimal
dust"~\cite{Eckhardt84, Henne08, Henne09, Kruger11_PRL, Kruger12_PRE, Kruger12}. The
infinitesimal absorption ensures that all electromagnetic waves eventually dissipate as
they propagate to infinity.  To make this explicit, we consider the environment as a
homogeneous background with a relative dielectric function $\epsilon_{\rm en}(\omega)$
that slightly deviates from unity~\cite{Kruger11_PRL, Kruger12_PRE, Kruger12}. The wave
equation in Eq.~\eqref{A_EOM} in such a medium is replaced by
\begin{equation} \label{A_EOM_eta}
  \epsilon_0 \epsilon_{\rm en} \partial_t^2 \bm{A} + \mu_0^{-1} \nabla \times \big(
  \nabla \times \bm{A} \big) = \bm{j} ,
\end{equation}
which yields
\begin{equation}
  \bm{A} = - D_0^r \bm{j} ,
\end{equation}
with
\begin{equation} \label{D0_en}
  D_0^r(\bm{Q},\omega) = 
\frac{\mathbb{I} - \bm{Q} \bm{Q}/(k_0^2 \epsilon_{\rm
  en})}{\epsilon_0 \big( \omega^2 \epsilon_{\rm en} - c^2 Q^2 \big)} .
\end{equation}
The equivalence between Eqs.~\eqref{D0_eta} and \eqref{D0_en} can be established through
the relation
\begin{equation}
  \epsilon_{\rm en}(\omega) = 1+ 2 i\eta / \omega + O(\eta^2) ,
\end{equation}
which corresponds to a medium with an infinitesimally small conductivity $\sigma = 2\eta
\epsilon_0$, i.e., $\epsilon_{\rm en}(\omega) = 1 + i\sigma(\omega) / (\epsilon_0\omega)$. 

This physical picture allows us to treat the environment's effect as an additional
self-energy within the Green's function formalism.
Modeling the dust as a homogeneous, isotropic, nonmagnetic medium renders the effective
self-energy $\Pi_0^r(\omega)$ local in space and independent of the wavevector $\bm{Q}$.
From Eqs.~\eqref{v_qw} and \eqref{D0_en}, one finds that $D_0^r$ satisfies
\begin{equation} \label{optical_0}
  \big[ D_0^r(\bm{Q},\omega)\big]^{-1} = 
  \big[ v(\bm{Q},\omega)\big]^{-1} - \Pi_0^r(\omega) ,
\end{equation}
which leads to the Dyson equation
\begin{equation} \label{Dyson_D0}
  D_0^r = v + v \Pi^r_0 D_0^r = v + D_0^r \Pi_0^r v ,
\end{equation}
where the retarded self-energy from the environment is
\begin{equation}
  \Pi^r_0(\omega) = -\epsilon_0 \omega^2 [\epsilon_{\rm en}(\omega) -1] 
  = -2 i\eta \epsilon_0 \omega  .
\end{equation}
In real space, this self-energy is local with
\begin{equation}
  \Pi^r_0(\bm{r},\bm{r}',\omega) = \Pi^r_0(\omega) \delta^3(\bm{r}-\bm{r}') .
\end{equation}
Consequently, the Dyson equation in real space takes the form of an integral equation
\begin{align}
  & D_0^r(\bm{r}-\bm{r}', \omega) = \notag \\
  & v(\bm{r}-\bm{r}', \omega) + \Pi^r_0(\omega) \int d^3\bm{r}_1 
    v(\bm{r}-\bm{r}_1, \omega)  D_0^r(\bm{r}_1 - \bm{r}', \omega) .
\end{align}
The free Green's function in real space is obtained by Fourier transforming
Eq.~\eqref{v_qw} with
\begin{equation}
  v(\bm{r},\omega)=\int \frac{d^3\bm{Q}}{(2\pi)^3} v(\bm{Q},\omega) e^{i\bm{Q}\cdot \bm{r}}
  = -\mu_0\Big(\mathbb{I} +\frac{1}{k_0^2} \nabla\nabla \Big) \frac{e^{ik_0 r}}{4\pi r} ,
\end{equation}
which evaluates to
\begin{align}
  & v(\bm{r}, \omega) =\notag \\
  & -\frac{\mu_0 e^{ik_0 r}}{4\pi r} \Big[\mathbb{I} - \hat{\bm{r}} \hat{\bm{r}} -
  \Big(\frac{1}{ik_0 r} + \frac{1}{k_0^2 r^2} \Big) (\mathbb{I} - 3 \hat{\bm{r}}
  \hat{\bm{r}})\Big] + \frac{\delta^3(\bm{r})}{3\epsilon_0 \omega^2} \mathbb{I},
  \label{v_rw}
\end{align}
with $r = |\bm{r}|$ and $\hat{\bm{r}} = \bm{r}/r$ the radial direction unit vector.
The retarded Green's function $D_0^r(\bm{r}, \omega)$ is obtained from $v(\bm{r}, \omega)$
by replacing $\omega$ with $\omega + i\eta$.
The advanced components are given by the Hermitian conjugate with
\begin{align}
  & D_0^a(\bm{Q},\omega) = \big[ D_0^r(\bm{Q},\omega) \big]^{\dag} , \\
  & D_0^a(\bm{r}-\bm{r}', \omega) = \big[ D_0^r(\bm{r}'-\bm{r}, \omega) \big]^{\dag}.
\end{align}

A fundamental relation follows from Eq.~\eqref{optical_0} by taking the difference between
the inverse retarded and advanced Green's functions, which in momentum space
yields~\cite{Henne08}
\begin{equation}
  \big[ D_0^a(\bm{Q}, \omega) \big]^{-1} -\big[ D_0^r(\bm{Q}, \omega) \big]^{-1}
  = \Pi_0^r(\omega) - \Pi_0^a(\omega).
\end{equation}
Fourier transforming back to real space and using the locality of the self-energy, we
obtain
\begin{align} 
  & \big[ D_0^a(\bm{r} - \bm{r}', \omega) \big]^{-1} - \big[ D_0^r(\bm{r} - \bm{r}',
  \omega) \big]^{-1} \notag \\
  = & [\Pi_0^r(\omega) - \Pi_0^a(\omega)] \delta^3(\bm{r} - \bm{r}'), \label{optical_1}
\end{align}
where the inverses are understood in the operator sense.
This relation will be useful to obtain the generalized optical theorem used later in this
work. Using the fact that $\Pi_0^r$ and $\Pi_0^a$ are purely imaginary, we arrive at
\begin{equation} 
  \Pi_0^r = \frac{1}{2} \Big[ \big(D_0^a \big)^{-1} - \big( D_0^r\big)^{-1} \Big] ,
\end{equation}
which holds both in real space and in momentum space.

As a thermal reservoir at temperature $T_0$, the environment naturally satisfies the
fluctuation-dissipation theorem.
The Keldysh Green's function $D_0^K(\omega)$ is therefore related to its retarded and
advanced components through~\cite{Agarwal75-1, Eckhardt84, Lifshitz_book}
\begin{equation} \label{D0K_1}
  D_0^K(\omega) = 2\big[D_0^r(\omega) - D_0^a(\omega) \big] [N_0(\omega) + 1/2] ,
\end{equation}
where the Bose-Einstein distribution function for the environment is 
\begin{equation}
  N_0(\omega)= 1/ \left\{ \exp[\hbar\omega/(k_B T_0)]-1 \right\} ,
\end{equation}
and the term $1/2$ accounts for the quantum vacuum fluctuations.
Combining Eq.~\eqref{optical_1} and \eqref{D0K_1} yields an equivalent and often more
convenient formulation, expressing the Keldysh Green's function directly in terms of the
self-energy with
\begin{equation} \label{D0K_2}
  D_0^K(\omega) = D_0^r(\omega) \Pi_0^K(\omega) D_0^a(\omega)  ,
\end{equation}
where the Keldysh self-energy is given by
\begin{equation} \label{Pi0K}
  \Pi_0^K(\omega) = 2\big[\Pi_0^r(\omega) - \Pi_0^a(\omega) \big] [N_0(\omega) + 1/2].
\end{equation}
Equations~\eqref{D0K_1} and \eqref{D0K_2} hold in both real and momentum space, with the
latter involving a convolution over spatial coordinates in the real-space representation.

Before proceeding, it is essential to distinguish the free-space vacuum Green's function
$v$ from the bare retarded one $D_0^r$. The latter obeys the fluctuation-dissipation
relation by virtue of the infinitesimal dissipation required to ensure causality.

\subsection{Green's function in the presence of an object}
We now construct the full photonic Green's function by incorporating the influence of the
material object through its self-energy.
The full photonic Green's function $D$ is related to the photonic self-energy $\Pi_1$ and
the bare Green's function $D_0$ by the Dyson equation on the Keldysh contour as
\begin{align} \label{Dyson_K}
  D &(\bm{r}, \tau; \bm{r}',\tau') = D_0(\bm{r},\tau; \bm{r}',\tau') + \iiiint d\tau_1
  d\tau_2 d^3\bm{r}_1 d^3\bm{r}_2  \notag \\
  & D_0(\bm{r},\tau;\bm{r}_1,\tau_1) \Pi_1(\bm{r}_1,\tau_1;\bm{r}_2,\tau_2)
  D(\bm{r}_2,\tau_2;\bm{r}',\tau').
\end{align}
From the Langreth's rules~\cite{Langreth, Haug_Jauho} and working in the frequency domain,
the retarded Green's function satisfies the Dyson equation
\begin{align} \label{Dyson_real}
  D^r(&\bm{r},\bm{r}') = D_0^r(\bm{r}-\bm{r}')  \notag \\
  +& \iint d^3\bm{r}_1 d^3\bm{r}_2
    D_0^r(\bm{r}-\bm{r}_1) \Pi^r_1(\bm{r}_1,\bm{r}_2) D^r(\bm{r}_2,\bm{r}') .
\end{align}
where the frequency argument $\omega$ is suppressed for brevity. 
The retarded self-energy $\Pi^r_1$ describes the linear response of the induced current to
the vector potential:
\begin{equation}
  \langle \bm{j}_{\rm ind}(\bm{r}) \rangle = - \int d^3\bm{r}' \Pi^r_1(\bm{r}, \bm{r}')
  \langle \bm{A}(\bm{r}') \rangle ,
\end{equation}
and is related to the dielectric function via~\cite{Lifshitz_book}
\begin{equation}
  \Pi^r_1(\bm{r},\bm{r}',\omega) = -\epsilon_0 \omega^2
  [\epsilon_1(\bm{r},\bm{r}',\omega) -1] .
\end{equation}
Unlike the environment self-energy which reduces to a local form via the Dirac delta
function due to vacuum homogeneity, the object's self-energy depends on two spatial
coordinates to account for possible inhomogeneity and spatial dispersion.
Equation~\eqref{Dyson_real} can be written compactly as
\begin{equation} \label{Dyson}
  D^r = D_0^r + D_0^r \Pi_1^r D^r 
  = D_0^r + D^r \Pi_1^r D_0^r ,
\end{equation}
or in differential form as
\begin{equation} \label{Dyson_diff}
  D^r(D_0^r)^{-1} = \mathbb{I} + D^r \Pi^r_1 ,
\end{equation}
where $\mathbb{I}$ denotes $\mathbb{I} \delta^3(\bm{r}-\bm{r}')$ in real space.
Recalling Eq.~\eqref{Dyson_D0}, an equivalent formulation that includes the environmental
self-energy $\Pi_0$ is
\begin{equation} \label{Dyson_1}
  D^r = v + v (\Pi_0^r + \Pi_1^r) D^r .
\end{equation}
The self-energy $\Pi_1$ is only nonzero within the object. Although $\Pi_0^r$ is
infinitesimally small and can be neglected inside the object, it plays an essential role
in the surrounding environment.
We denote by $V_0$ the spatial region outside the object, representing the environment.
The physical significance of this term is that it accounts for photons that escape to
infinity, causing the environment to act as an additional dissipative bath.
Because the self-energy $\Pi_0$ and subsequently the bath at infinity $\Pi_\infty$
characterize the surrounding vacuum, their forms are invariant to the physical properties
of the object. Furthermore, the bath acts exclusively as a sink for propagating far-field
modes, as evanescent waves decay exponentially with distance and cannot reach infinity.

From the Dyson equation in Eq.~\eqref{Dyson_1}, we obtain the following relation:
\begin{equation} \label{optical_22}
  \big(D^a\big)^{-1} - \big( D^r \big)^{-1} = \big( \Pi_0^r - \Pi_0^a\big) + \big( \Pi_1^r
  - \Pi_1^a \big) .
\end{equation}
This identity leads to the generalized optical theorem for the spectral function
with~\cite{Eckhardt84, Henne08}
\begin{equation} \label{optical_2}
  D^r - D^a = X_{\rm en} + D^r \big(\Pi_1^r - \Pi_1^a \big) D^a,
\end{equation}
where the contribution from the environment is
\begin{align} 
  X_{\rm en} 
  &= D^r \big(\Pi_0^r -\Pi_0^a\big) D^a \notag \\
  &= \big(\mathbb{I} + D^r \Pi^r_1 \big) D_0^r (\Pi_0^r - \Pi_0^a) D_0^a
  \big(\mathbb{I} + \Pi^a_1 D^a \big) .
\label{X_en}
\end{align}
This has been obtained in Ref.~\cite{Eckhardt84} but not evaluated. 
The term $X_{\rm en}$ involves
\begin{align} 
  \Xi(\bm{r}, \bm{r}') =& \iint_{V_0} d^3\bm{r}_i d^3\bm{r}_j
  D_0^r(\bm{r} - \bm{r}_i) \big[\Pi_0^r(\bm{r}_i, \bm{r}_j)  \notag \\
  &\qquad \qquad \qquad - \Pi_0^a(\bm{r}_i, \bm{r}_j)\big] D_0^a(\bm{r}_j - \bm{r}')
  \notag \\
  =& -4 i\eta \epsilon_0 \omega \int_{V_0} d^3\bm{r}_i D_0^r(\bm{r} - \bm{r}_i)
  D_0^a(\bm{r}_i - \bm{r}') .  \label{Xi} 
\end{align}
The integration is restricted to the region $V_0$ outside the object.
This contribution from the infinitesimal environmental dust is nontrivial since
integrating such infinitesimal dust over the infinite space of the environment yields a
finite and physically meaningful result.

Using the Langreth's rules to Eq.~\eqref{Dyson_K}, the Keldysh component of the Green's
function takes the form~\cite{Keldysh64, Haug_Jauho}
\begin{equation}
  D^K = D_0^K + D_0^K \Pi^a_1 D^a + D_0^r \Pi^K_1 D^a + D_0^r \Pi^r_1 D^K ,
\end{equation}
where spatial coordinates and convolutions are suppressed for brevity.
Using Eqs.~\eqref{Dyson} and \eqref{Dyson_diff}, this expression simplifies to
\begin{equation} \label{DK}
  D^K = D^K_{\rm en} + D^r \Pi^K_1 D^a ,
\end{equation}
with the environmental contribution given by
\begin{equation}
  D^K_{\rm en} = \big(\mathbb{I} + D^r \Pi^r_1 \big) D_0^K \big(\mathbb{I} + \Pi^a_1 D^a
  \big) .
\end{equation}
Using Eqs.~\eqref{D0K_2} and \eqref{Dyson}, we obtain
\begin{equation} \label{DK_en}
  D^K_{\rm en} 
  = D^r \Pi^K_0 D^a = 2 X_{\rm en} [N_0(\omega) + 1/2] .
\end{equation}
Evaluating $X_{\rm en}$, which requires computing the integral in Eq.~\eqref{Xi}, is
therefore essential for determining the photonic distribution function.

An alternative derivation of Eq.~\eqref{DK} follows directly from the Dyson equation on
the Keldysh contour,
\begin{equation}
  D = v + v (\Pi_0 + \Pi_1) D .
\end{equation}
Applying the Langreth's rules and using the fact that the Keldysh component of the free
vacuum Green's function $v^K$ vanishes, we obtain
\begin{equation}
  D^K = v \Pi^K D^a + v \Pi^r D^K .
\end{equation}
Combining this with Eq.~\eqref{Dyson_1} then yields Eq.~\eqref{DK}.

In the following two sections, we evaluate the integrals in Eq.~\eqref{Xi} for two
representative scenarios: a finite three-dimensional object [Fig.~\ref{fig1}(a)] and a
semi-infinite system with in-plane translational invariance [Fig.~\ref{fig1}(b)].

\section{The three-dimensional photonic bath at infinity} \label{SecIII}
At asymptotically large distances with $r \gg 1/k_0 = c /\omega$, the higher-order terms
in $1/r$ become negligible. Retaining only the leading term, the free Green's functions in
Eq.~\eqref{v_rw} simplify to the far-field form
\begin{equation}
  v(\bm{r}, \omega) = -\frac{\mu_0}{4\pi}\frac{e^{ik_0 r}}{r} (\mathbb{I} - \hat{\bm{r}}
  \hat{\bm{r}}) .
\end{equation}

For the integral in Eq.~\eqref{Xi}, we can restrict the integration to the region with
$|\bm{r} - \bm{r}_i| \gg 1/k_0$, and $|\bm{r}_i - \bm{r}'| \gg 1/k_0$.
This simplification holds for any finite $\bm{r}$, $\bm{r}'$.
Due to the spherical symmetry, we adopt spherical coordinates centered at the origin.
The volume integral over the region outside a sphere of radius $R$, which is chosen
sufficiently large to satisfy the far-field condition, then becomes 
\begin{equation} 
  \Xi(\bm{r}, \bm{r}') = -4 i\eta \epsilon_0 \omega \int_R^{\infty} r_i^2 dr_i \int
  d\hat{\bm{r}}_i D_0^r(\bm{r} - \bm{r}_i) D_0^a(\bm{r}_i - \bm{r}') .
\end{equation}
To proceed, we employ the far-field approximations for the Green's functions.
For large $r_i$, we have
\begin{align}
  D_0^r(\bm{r} - \bm{r}_i) &\approx - \frac{\mu_0}{4\pi} \frac{e^{i k_0 |\bm{r} -
  \bm{r}_i|}}{|\bm{r} - \bm{r}_i|} e^{-\eta r_i/c} (\mathbb{I} - \hat{\bm{r}}_i
  \hat{\bm{r}}_i) , \label{D0rf} \\
  D_0^a(\bm{r}_i - \bm{r}') &\approx - \frac{\mu_0}{4\pi} \frac{e^{-i k_0 |\bm{r}_i -
  \bm{r}'|}}{|\bm{r}_i - \bm{r}'|} e^{-\eta r_i/c} (\mathbb{I} - \hat{\bm{r}}_i
  \hat{\bm{r}}_i) ,
  \label{D0af}
\end{align}
where we have used the fact that the unit vectors of $\bm{r}-\bm{r}_i$ and
$\bm{r}_i-\bm{r}'$ both approach $\hat{\bm{r}}_i$ for large $r_i$.
Furthermore, we approximate the amplitudes as 
$1/|\bm{r} - \bm{r}_i| \approx 1/|\bm{r}_i - \bm{r}'| \approx 1/r_i$.
The phases in Eqs.~\eqref{D0rf} and \eqref{D0af} can be treated by
\begin{equation}
  |\bm{r} - \bm{r}_i| \approx  r_i - \hat{\bm{r}}_i \cdot \bm{r} , \quad
  |\bm{r}_i - \bm{r}'| \approx r_i - \hat{\bm{r}}_i \cdot \bm{r}' .
\end{equation}
Substituting these approximations and using the projector property 
\begin{equation} \label{project_3D}
  (\mathbb{I} - \hat{\bm{r}}_i \hat{\bm{r}}_i)^2 = \mathbb{I} - \hat{\bm{r}}_i
  \hat{\bm{r}}_i , 
\end{equation}
we obtain
\begin{align}
  \Xi(\bm{r}, \bm{r}') = -4 i\eta \epsilon_0 \omega \Big(\frac{\mu_0}{4\pi} \Big)^2 
  & \int_R^{\infty} dr_i e^{-2\eta r_i/c} \int d\hat{\bm{r}}_i   \notag \\
  & e^{i k_0 (\bm{r}' - \bm{r}) \cdot \hat{\bm{r}}_i} (\mathbb{I} - \hat{\bm{r}}_i
  \hat{\bm{r}}_i) .
\end{align}
In the limit $\eta \to 0^+$, we have the radial integral
\begin{equation}
  \eta \int_R^\infty dr_i e^{-2\eta r_i/c} = \frac{c}{2}e^{-2\eta R/c} \to \frac{c}{2} .
\end{equation}
Thus, one has
\begin{equation} \label{Xi_angular}
  \Xi(\bm{r}, \bm{r}') = -2 i \epsilon_0 c \omega \Big(\frac{\mu_0}{4\pi} \Big)^2 \int
  d\Omega e^{i k_0 (\bm{r}' - \bm{r}) \cdot \hat{\bm{R}}} (\mathbb{I} - \hat{\bm{R}}
  \hat{\bm{R}}) ,
\end{equation}
where we have changed the dummy angular integration variable from $\hat{\bm{r}}_i$ to unit
vector $\hat{\bm{R}} = \bm{R}/R$ and denote the differential solid angle element as
$d\Omega$.
The angular integral in Eq.~\eqref{Xi_angular} is a function of $\bm{r}$ and $\bm{r}'$.
To connect with the concept of a bath defined on a spherical surface, we use the far-field
approximations for the bare Green's functions with
\begin{align}
  D_0^r(\bm{r} - \bm{R}) &\approx - \frac{\mu_0}{4\pi} \frac{e^{i k_0 (R -
  \hat{\bm{R}}\cdot \bm{r})}}{R} (\mathbb{I} - \hat{\bm{R}} \hat{\bm{R}}) , 
  \label{D0r_R}\\
  D_0^a(\bm{R} - \bm{r}') &\approx - \frac{\mu_0}{4\pi} \frac{e^{-i k_0 (R -
  \hat{\bm{R}}\cdot \bm{r}')}}{R} (\mathbb{I} - \hat{\bm{R}} \hat{\bm{R}}) ,
  \label{D0a_R}
\end{align}
where the infinitesimal $\eta$ has been dropped since the integration over the infinite
region has already been carried out.
We can arrive at a mathematically equivalent equation to Eq.~\eqref{Xi_angular} with 
\begin{equation} \label{Xi_infty_3D}
  \Xi(\bm{r},\bm{r}') = \int_S D_0^r(\bm{r}-\bm{R}) [\Pi^r_\infty(\bm{R}) -
  \Pi^a_\infty(\bm{R})] D_0^a(\bm{R}-\bm{r}'),
\end{equation}
with the shorthand notation
\begin{equation}
  \int_S \equiv \int R^2d\Omega .
\end{equation}
In this step, we have identified the effective self-energy of the photonic bath at
infinity as
\begin{equation} \label{Pi_infty_3D}
  \Pi_{\infty}^r(\bm{R}) = \big[ \Pi_{\infty}^a(\bm{R})\big]^* = 
  -i\epsilon_0 c \omega \big( \mathbb{I} - \hat{\bm{R}}\hat{\bm{R}} \big) .
\end{equation}
In Eq.~\eqref{Xi_infty_3D}, the factor $R^2$ from the surface element is exactly canceled
by the $1/R^2$ dependence in the product of the two far-field Green's functions.
As a result, the entire expression is independent of the chosen radius $R$, as required
for a physically meaningful bath.
With this result, the environmental contribution in Eq.~\eqref{X_en} can be simplified to
\begin{equation} \label{X_en_infty}
  X_{\rm en}(\bm{r}, \bm{r}') = \int_S D^r(\bm{r}, \bm{R}) 
  [\Pi^r_\infty(\bm{R}) - \Pi^a_\infty(\bm{R})] D^a(\bm{R}, \bm{r}') ,
\end{equation}
where the bare Green's functions $D_0^{r/a}$ in Eq.~\eqref{Xi_infty_3D} have been replaced
by the full Green's functions $D^{r/a}$ to account for the presence of the object.
For clarity, we reiterate that $\Pi_0$ characterizes the self-energy of a homogeneous
environment filling the entire space, whereas $\Pi_\infty$ denotes the effective
self-energy of the photonic bath at infinity on a sphere.

One might be tempted to evaluate the environment's contribution to Eq.~\eqref{Dyson_1}
directly and obtain the effective self-energy through the integral
\begin{equation} 
   -2 i\eta \epsilon_0 \omega \int_{V_0} d^3\bm{r}_i v(\bm{r} -
  \bm{r}_i) D_0^r(\bm{r}_i - \bm{r}') .
\end{equation}
However, this approach fails because the phase factors in $v(\bm{r} - \bm{r}_i)$ and
$D_0^r(\bm{r}_i - \bm{r}')$ both depend on $k_0 r_i$ in the far-field region, resulting in
a net phase factor $e^{2ik_0 r_i}$ that oscillates rapidly as $r_i \to \infty$.
Consequently, the integral over the infinite volume $V_0$ is not convergent.

Several remarks are in order regarding Eq.~\eqref{Pi_infty_3D}, which is the first central
result of this work.
First, the self-energy is purely imaginary, reflecting the dissipative nature of the
photonic bath.
Second, we have explicitly attached the transverse projector to $\Pi_{\infty}^r$ since
those in the bare Green's functions $D_0^r(\bm{r} - \bm{R})$ and $D_0^a(\bm{R} - \bm{r}')$
can be absorbed or modified by the object's self-energy when simplifying Eq.~\eqref{X_en}
to get Eq.~\eqref{X_en_infty}.
Third, the self-energy can be expressed as $\Pi_{\infty}^r(\omega) = - i (\omega / Z_0)
\big(\mathbb{I} -\hat{\bm{R}}\hat{\bm{R}}\big)$, where $Z_0 = \sqrt{\mu_0/\epsilon_0}$ is
the vacuum impedance.
This linear scaling with frequency demonstrates that in three dimensions the bath at
infinity acts as a purely Ohmic environment~\cite{Weiss}.
Finally, in the static limit $\omega \to 0$, the diverging length scale $1/k_0$ pushes the
effective boundary of the environment to infinity, physically justifying the nomenclature
``photonic bath at infinity".

\subsection{Bath at infinity behaves as a black body}
The energy density $u$ at the origin, $\bm{r}=(0,0,0)$, due to the photonic bath at
infinity at temperature $T_0$ should be given by the Planck formula, as such a bath
behaves as a perfect black body. This has been discussed in Ref.~\cite{JSW22}, and we
include the following derivation to make our presentation self-contained.

The electromagnetic energy density is defined using the normally ordered field operators
to exclude the divergent vacuum zero-point 
contribution with
\begin{equation}
  u(\bm{r}) = \frac{1}{2} \bigg[ \epsilon_0 \langle : \bm{E}^2(\bm{r},t) : \rangle +
  \frac{1}{\mu_0} \langle : \bm{B}^2(\bm{r},t) : \rangle \bigg].
\end{equation}
Expressing the fields via Eq.~\eqref{EABA} allows the energy density to be cast in terms
of photonic Green's functions, where normal ordering specifically corresponds to the use
of the lesser Green's function $D^<$. Transforming to the frequency domain, we obtain
\begin{align} \label{u_D}
 u = \frac{1}{2} \int_0^\infty & \frac{d\omega}{\pi} 
  i\hbar \, {\rm tr} \bigg[ \epsilon_0 \omega^2 D_0^<(\bm{r}, \bm{r}') \notag \\
  &+ \frac{1}{\mu_0} \nabla_{\bm{r}} \times D_0^<(\bm{r}, \bm{r}') \times 
  \overleftarrow{\nabla}_{\bm{r}'} \bigg]_{\bm{r}=\bm{r}'=\bm{0}} .
\end{align}
with the definition
\begin{equation}
  \big[ \nabla_{\bm{r}} \times F(\bm{r}, \bm{r}') \times \overleftarrow{\nabla}_{\bm{r}'}
  \big]_{\mu\nu} \equiv \varepsilon_{\mu\alpha\beta} \varepsilon_{\nu\gamma\delta}
  \partial_{\alpha} \partial'_{\gamma} F_{\beta\delta}(\bm{r},\bm{r}') ,
\end{equation}
where $\partial_{\alpha} = \partial / \partial \bm{r}_{\alpha}$, 
$\partial'_{\gamma} = \partial / \partial \bm{r}'_{\gamma}$, $\varepsilon$ is the
Levi-Civita symbol, and the summation over repeated indices is implied.

The lesser component of the bare Green's function satisfies the Keldysh equation
\begin{equation} \label{D0K}
  D_0^<(\bm{r}, \bm{r}') = \int R^2 d\Omega D_0^r(\bm{r}, \bm{R}) \Pi^<_\infty(\bm{R})
  D_0^a(\bm{R}, \bm{r}') ,
\end{equation}
where the bath's lesser self-energy is given by the fluctuation–dissipation relation
\begin{equation} \label{Pi<_infty_FD}
  \Pi^<_\infty(\omega) = [\Pi^r_\infty(\omega) - \Pi^a_\infty(\omega)] N_0(\omega) .
\end{equation}
Together with the far-field approximations for $D_0^r$ and $D_0^a$ in Eqs.~\eqref{D0r_R}
and \eqref{D0a_R}, and using 
\begin{equation}
  {\rm tr}(\mathbb{I} - \hat{\bm{R}} \hat{\bm{R}})=2 , 
\end{equation}
we find 
\begin{equation} 
  {\rm tr}\big[D_0^<(\bm{r},\bm{r})\big] =\frac{-i\omega}{\pi \epsilon_0 c^3} N_0(\omega).
\end{equation}

For the magnetic contribution, we use the general identity for a dyadic field 
$F(\bm{r}, \bm{r}')$ with
\begin{equation} \label{curlcurl}
  {\rm tr} \big[ \nabla_{\bm{r}} \times F(\bm{r}, \bm{r}') \times
  \overleftarrow{\nabla}_{\bm{r}'} \big] 
  = \nabla_{\bm{r}} \cdot \nabla_{\bm{r}'}\, {\rm tr} (F) - \nabla_{\bm{r}} \cdot
  (\nabla_{\bm{r}'} \cdot F) ,
\end{equation}
where we have used the relation 
$\varepsilon_{\mu\alpha\beta} \varepsilon_{\mu\gamma\delta} =
\delta_{\alpha\gamma}\delta_{\beta\delta} - \delta_{\alpha\delta}\delta_{\beta\gamma}$, 
and the notation for the inner divergence as
$(\nabla_{\bm{r}'} \cdot F)_{\alpha}\equiv \partial'_{\beta}F_{\beta\alpha}$.
For free-space radiation, the lesser Green's function is transverse with
$\nabla_{\bm{r}'} \cdot D_0^<(\bm{r}- \bm{r}')=0$.
Moreover, due to translational invariance, $D_0^<$ depends only on $\bm{r}-\bm{r}'$,
implying $\nabla_{\bm{r}'} = - \nabla_{\bm{r}}$. Equation~\eqref{curlcurl} then reduces to
\begin{align}
  {\rm tr} \big[ \nabla_{\bm{r}} \times D_0^<(\bm{r}- \bm{r}')
  \times \overleftarrow{\nabla}_{\bm{r}'} \big] 
  &= -\nabla_{\bm{r}}^2 {\rm tr} \big[ D_0^<(\bm{r}- \bm{r}') \big] \notag \\
  &= k_0^2 {\rm tr} \big[ D_0^<(\bm{r} - \bm{r}) \big] .
\end{align}
Summing the equal contributions from the electric and magnetic fields, we obtain the total
energy density:
\begin{equation} \label{Planck}
  u = \int_0^\infty d\omega \frac{\hbar\omega^3}{\pi^2 c^3} N_0(\omega)  
  = \frac{4 \sigma_{\rm SB}}{c} T^4 ,
\end{equation}
where $\sigma_{\rm SB} = \pi^2 k_B^4 / (60 c^2 \hbar^3)$ is the Stefan-Boltzmann constant.
This result is the standard expression for the energy density of blackbody radiation at
temperature $T$.

\subsection{Far-field thermal radiation}
From the perspective of Joule heating, the heat current flowing out of the object is
expressed as
\begin{equation}
  J(t) = -\int_{V_1} d^3\bm{r} \left< \bm{j} \cdot \bm{E} \right> ,
\end{equation}
where $V_1$ denotes the spatial region of the object.
Using Eq.~\eqref{A_EOM} and the relation $\bm{E}=-\partial \bm{A}/\partial t$, the heat
current can be expressed in terms of the photonic Green's function as
\begin{align} \label{Jt}
  J &= -\int_{V_1} d^3\bm{r} \bigg[ \frac{\partial\ }{\partial t'} \sum_{\mu\nu}
  v^{-1}_{\mu\nu}(\bm{r}, t) \big< A_\nu(\bm{r}, t) A_\mu(\bm{r}, t') \big>
  \bigg]_{t'=t}  \notag \\
  &= -\frac{i\hbar}{2}\int_{V_1} d^3\bm{r} \bigg\{ \frac{\partial\ }{\partial t'} {\rm tr} 
  \big[ v^{-1}(\bm{r}, t) D^K(\bm{r}, t; \bm{r}, t')\big] \bigg\}_{t'=t} ,
\end{align}
where the trace is taken over the Cartesian components.
Here, $v^{-1}$ represents the inverse photon propagator in its differential operator form,
while the time derivative $\partial/\partial t'$ acts on the second temporal argument of
$D^K$. The use of the Keldysh Green's function is essential because the heat current as a
physical observable must be real. While the product of two arbitrary Hermitian operators
is generally not Hermitian, the symmetrized combination yields a real expectation value.

Using the Keldysh equation in Eq.~\eqref{DK} and the Dyson equation in Eq.~\eqref{Dyson_1}
in the differential form $v^{-1} D^r = \mathbb{I} + (\Pi_0^r + \Pi_1^r) D^r$, we obtain
$v^{-1} D^K = (\Pi_0^K + \Pi_1^K) D^a + (\Pi_0^r + \Pi_1^r) D^K$.
Putting this result into Eq.~\eqref{Jt}, we obtain the Meir-Wingreen formula for the heat
current~\cite{JSW23}. In frequency space, it can be expressed as
\begin{align} 
  J = \int_{-\infty}^{\infty} \frac{d\omega}{4\pi} \hbar\omega \iint_{V_1} d^3\bm{r} 
  & d^3\bm{r}' {\rm tr} \big[ \Pi_1^K(\bm{r}, \bm{r}')D^a(\bm{r}', \bm{r}) \notag \\
  &+ \Pi_1^r(\bm{r}, \bm{r}') D^K(\bm{r}', \bm{r}) \big] .
\end{align}
Using Eqs.~\eqref{symD_21} and \eqref{symD_22}, we have
\begin{align} 
  J = & \int_0^{\infty} \frac{d\omega}{4\pi} \hbar\omega \iint_{V_1} d^3\bm{r} 
  d^3\bm{r}' {\rm tr} \big\{ \Pi_1^K(\bm{r}', \bm{r}) \big[ D^a(\bm{r}, \bm{r}') 
  \notag \\
  &- D^r(\bm{r}, \bm{r}') \big] + \big[ \Pi_1^r(\bm{r}', \bm{r}) - \Pi_1^a(\bm{r}',
  \bm{r})\big] D^K(\bm{r}, \bm{r}') \big\} .
\end{align}
Under local thermal equilibrium, the fluctuation-dissipation relation tells that
\begin{equation}
  \Pi^K_1 = 2\ \big(\Pi^r_1 - \Pi^a_1 \big) (N_1 + 1/2) 
  = 4i\ {\rm Im}\big(\Pi^r_1\big) (N_1 + 1/2) ,
\end{equation}
with the Bose-Einstein distribution function for the object as
\begin{equation}
  N_1(\omega)= 1/ \left\{ \exp[\hbar\omega/(k_B T_1)]-1 \right\} .
\end{equation}
Using Eqs.~\eqref{optical_2}, \eqref{DK} and \eqref{DK_en}, we obtain
\begin{align} 
  J = -2i \int_0^{\infty} \frac{d\omega}{2\pi} \hbar\omega 
  & \iint_{V_1} d^3\bm{r} d^3\bm{r}' {\rm tr} \big\{ {\rm Im}
    \big[\Pi_1^r(\bm{r}', \bm{r}) \big]  \notag \\
  & X_{\rm en}(\bm{r}, \bm{r}') [N_1(\omega) - N_0(\omega)] \big\} , \label{JJ}
\end{align}
where $X_{\rm en}(\bm{r}, \bm{r}')$ is given in Eq.~\eqref{X_en_infty}.
We therefore can arrive at a Landauer-B\"{u}ttiker form as~\cite{JSW23, Kiryl25}
\begin{equation} 
  J = \int_0^{\infty} \frac{d\omega}{2\pi} \hbar\omega 
  \iint_{V_1} d^3\bm{r} d^3\bm{r}'
  \xi(\bm{r}, \bm{r}', \omega) [N_1(\omega) - N_0(\omega)] ,
\end{equation}
where the photonic transmission function $\xi$ is given by
\begin{align} 
  & \xi(\bm{r}, \bm{r}', \omega) = \notag \\
  & 4 \int_S {\rm tr} \big\{ {\rm Im} \big[\Pi_1^r(\bm{r}', \bm{r}) \big]  D^r(\bm{r},
    \bm{R}) {\rm Im} [\Pi^r_\infty(\bm{R})] D^a(\bm{R}, \bm{r}') \big\} ,
\end{align}
with integrating over the surface of the sphere and summing over the direction index. 

Under the local approximation, the self-energy of the object is given by
\begin{equation}
  \Pi_1^r(\bm{r}',\bm{r},\omega) = -\epsilon_0 
\omega^2 [\epsilon_1(\omega) -\mathbb{I}]
  \delta^3(\bm{r}-\bm{r}') ,
\end{equation}
which allows us to express the heat current as
\begin{align} 
  J = & -4 \epsilon_0 \int_0^{\infty} \frac{d\omega}{2\pi} \hbar\omega^3 
  [N_1(\omega) - N_0(\omega)] \int_{V_1} d^3\bm{r}  \notag \\
  & \int R^2 d\Omega \, {\rm tr} \big\{ {\rm Im} (\epsilon_1) D^r(\bm{r}, \bm{R})
  {\rm Im} [\Pi^r_\infty(\bm{R})] D^a(\bm{R}, \bm{r}) \big\} .
\end{align}
In the far-field limit where $R$ is much larger than the object's dimensions, we apply a
monopole approximation by replacing the Green's functions $D^r(\bm{r}, \bm{R})$ and
$D^a(\bm{R}, \bm{r})$ with the corresponding bare ones in Eqs.~\eqref{D0r_R} and
\eqref{D0a_R}.
Using the projector property in Eq.~\eqref{project_3D} and the identity
\begin{equation}
  \int \frac{d\Omega}{4\pi} \hat{\bm{R}} \hat{\bm{R}} = \frac{\mathbb{I}}{3} ,
\end{equation}
which can be verified by taking trace on both sides, we obtain~\cite{Kiryl17}
\begin{equation} \label{J_V1}
  J =  V_1\int_0^{\infty} d\omega \frac{\hbar\omega^4 }{3\pi^2 c^3} 
  {\rm tr} \{{\rm Im}[\epsilon_1(\omega)]\} [N_1(\omega) - N_0(\omega)] .
\end{equation}
In this optically thin limit, the object's self-emission is entirely equivalent to
the heat transfer from the object to a zero-temperature environment.
It is instructive to compare this result with the classical Stefan-Boltzmann law,
$J_{\rm SB} = \sigma_{\rm SB} A T^4$, where the heat current scales with surface area $A$.
This law describes an optically thick (blackbody) regime where the absorption length is
much smaller than the characteristic scale of the object, causing interior radiation to be
re-absorbed before reaching the boundary.
In contrast, Eq.~\eqref{J_V1} describes the optically thin limit with $\epsilon_1$ being
close to unity, which is representative of nanoparticles, thin films, or weakly absorbing
media. In this regime, the material is effectively transparent to its own thermal
fluctuations.
The absence of re-absorption allows each internal dipole to radiate independently,
resulting in a total power proportional to the volume.
Equation~\eqref{J_V1} is consistent with derivations based on the surface integral of the
Poynting vector~\cite{JSW20, JSW22, JSW23}.

\section{The two-dimensional photonic bath at infinity} \label{SecIV}
We now consider the effective self-energy for the system in Fig.~\ref{fig1}(b), which has
in-plane translational invariance and a planar interface normal to the $z$-axis.
It is convenient to work in the mixed representation (frequency $\omega$, in-plane
wavevector $\bm{q}$, and coordinate $z$). The environment's contribution $X_{\rm en}$ in
Eq.~\eqref{X_en} involves
\begin{equation} \label{Xi_2D_start} 
  \Xi(z, z') = -4 i\eta \epsilon_0 \omega \int_0^{\infty} dz_i D_0^r(z - z_i) D_0^a(z_i -
  z') ,
\end{equation}
where the arguments $\omega$ and $\bm{q}$ are suppressed for brevity.
The free-space Green's function in the mixed representation is obtained by Fourier
transforming Eq.~\eqref{v_qw} with~\cite{Sipe87} 
\begin{equation}
  v(\omega, \bm{q}, z) = \int \frac{d q_z}{2\pi} v(\omega, \bm{Q}) e^{iq_z z} 
  = \tilde{v}(\omega, \bm{q}, z) + \frac{\mu_0}{k_0^2}\delta(z) \hat{\bm{z}}\hat{\bm{z}},
\end{equation}
where $\hat{\bm{z}}$ is the unit vector in the $z$ direction and
\begin{equation}
  \tilde{v}(\omega,\bm{q},z)
  = \Upsilon \big[\hat{\bm{s}} \hat{\bm{s}} + \theta(z) \hat{\bm{p}}_+
  \hat{\bm{p}}_+ + \theta(-z) \hat{\bm{p}}_- \hat{\bm{p}}_-\big] e^{i\gamma_0|z|},
\end{equation}
with 
\begin{equation}
  \Upsilon = \frac{\mu_0}{2i\gamma_0}, \qquad  \gamma_0 = \sqrt{k_0^2 - q^2} .
\end{equation}
The polarization vectors for $s$ and $p$ modes are given by~\cite{Sipe87}
\begin{equation}
  \hat{\bm{s}} = \hat{\bm{q}}\times \hat{\bm{z}}, \qquad
  \hat{\bm{p}}_{\pm} =(\mp \gamma_0 \hat{\bm{q}} + q\hat{\bm{z}})/k_0 ,
\end{equation}
where the hat denotes a unit vector.
The $s$-polarization vector $\hat{\bm{s}}$ is
transverse to both the propagation direction and the $z$-axis, while the $p$-polarization
vectors $\hat{\bm{p}}_+$ and $\hat{\bm{p}}_-$ correspond to waves propagating in the $+z$
and $-z$ directions, respectively.
Their explicit component forms are
\begin{equation} \label{sp}
  \hat{\bm{s}} = \frac{1}{q} 
  \begin{bmatrix}
    q_y \\  -q_x \\ 0
  \end{bmatrix} , \qquad
  \hat{\bm{p}}_{\pm} = 
  \frac{1}{qk_0} 
  \begin{bmatrix}
    \mp\gamma_0 q_x \\  \mp\gamma_0 q_y \\ q^2
  \end{bmatrix} .
\end{equation}

For the integral in Eq.~\eqref{Xi_2D_start}, we restrict the integration to the region
$z_i > d$ with $d$ chosen sufficiently large (greater than both $z$ and $z'$).
This simplification holds for any finite $z$, $z'$. Therefore, we have
\begin{equation} \label{Xi_2D_far} 
  \Xi(z, z')= -4 i\eta \epsilon_0 \omega \int_d^\infty dz_i D_0^r(z - z_i) D_0^a(z_i -z').
\end{equation}
In the far-field region with $q<k_0$ and $z_i > z, z'$, 
using 
\begin{equation} \label{tilde_gamma_0}
  \tilde{\gamma}_0 =\sqrt{(\omega + i\eta)^2 /c^2 - q^2} =\gamma_0 +i\eta k_0 /c\gamma_0 ,
\end{equation}
the retarded and advanced Green's functions take the asymptotic forms that include the
infinitesimal absorption $\eta$ with
\begin{align}
  D_0^r(z - z_i) &\approx \Upsilon e^{i\gamma_0 (z_i- z)}e^{-\eta k_0 (z_i - z) /
  c\gamma_0} \mathbb{P} , \\
  D_0^a(z_i - z') &\approx -\Upsilon e^{i\gamma_0 (z'- z_i)}e^{-\eta k_0 (z_i - z') /
  c\gamma_0} \mathbb{P} ,
\end{align}
where we have introduced the shorthand
\begin{equation}
  \mathbb{P} = \hat{\bm{s}} \hat{\bm{s}} + \hat{\bm{p}}_- \hat{\bm{p}}_- ,
\end{equation}
and used the symmetry relation in Eq.~\eqref{symD_q11}.
Substituting these approximations into Eq.~\eqref{Xi_2D_far} and applying the projector
property $\mathbb{P}\mathbb{P}=\mathbb{P}$, we obtain
\begin{equation}
  \Xi(z, z') = 4 i\eta \epsilon_0 \omega \Upsilon^2 \int_d^{\infty} dz_i 
  e^{i\gamma_0(z' - z)} e^{2\eta k_0(z+z'-z_i)/c\gamma_0} \mathbb{P} .
\end{equation}
In the limit $\eta\to 0^+$, the integral over $z_i$ gives
\begin{equation}
  \eta \int_d^{\infty} dz_i e^{2\eta k_0 (z+z'-z_i)/c\gamma_0} = \frac{c\gamma_0}{2k_0} ,
\end{equation}
which upon substitution yields
\begin{equation}
  \Xi(z, z') = 2 i \epsilon_0 c^2 \gamma_0 \Upsilon^2 e^{i\gamma_0 (z' - z)} \mathbb{P} .
\end{equation}

Similarly to the three-dimensional case, to connect with the concept of a bath at
infinity, we employ the approximations for the far-field bare Green's functions with
\begin{equation}\label{D0f_2D} 
  D_0^r(z - d) \approx \Upsilon e^{i\gamma_0 (d- z)} \mathbb{P} , \quad
  D_0^a(d - z') \approx -\Upsilon e^{i\gamma_0 (z'- d)} \mathbb{P} ,
\end{equation}
where the infinitesimal $\eta$ has been dropped since the integration over the infinite
region has already been carried out.
For any reference plane $z_{\infty} \geq d$, we rewrite the result as
\begin{equation} \label{Xi_2D_surface}
  \Xi(\bm{q}, z, z') = D_0^r(\bm{q}, z, z_\infty) \big[ \Pi_\infty^r(\bm{q}) -
  \Pi_\infty^a(\bm{q}) \big] D_0^a(\bm{q}, z_\infty, z'),
\end{equation}
and identify the effective self-energy of the photonic bath at infinity:
\begin{equation} \label{Pi_2D_infty}
  \Pi_{\infty}^r(\bm{q}) = \big[ \Pi_{\infty}^a(\bm{q})\big]^* = 
  -i \epsilon_0 c^2 \gamma_0 P_{\parallel} .
\end{equation}
with the in-plane projection matrix $P_\parallel$ still to be determined.

Since the bath at infinity has no out-of-plane component, the projection matrix
$P_\parallel$ can be expressed as a combination of in-plane polarization vectors with
\begin{equation}
  P_\parallel = \hat{\bm{s}} \hat{\bm{s}} + \zeta \hat{\bm{p}} \hat{\bm{p}} ,
\end{equation}
where $\hat{\bm{p}}$ is the reduced $p$-polarization vector:
\begin{equation}
  \hat{\bm{p}} = 
  \big[\hat{\bm{p}}_{-}\big]_{\parallel} = 
  \frac{\gamma_0}{qk_0} 
  \begin{bmatrix}
    q_x \\ q_y \\ 0
  \end{bmatrix} ,
\end{equation}
and the scalar $\zeta$ is fixed by the consistency condition
\begin{equation} \label{PPP}
  \mathbb{P} \, P_\parallel \, \mathbb{P} = \mathbb{P} .
\end{equation}
While this condition permits infinite solutions, the absence of out-of-plane components
and the decoupling of $s$- and $p$-modes eliminate all arbitrary degrees of freedom,
uniquely determining $P_\parallel$. Solving Eq.~\eqref{PPP} yields 
\begin{equation}
  P_\parallel
  = \hat{\bm{s}} \hat{\bm{s}} + (k_0 / \gamma_0)^4 \hat{\bm{p}} \hat{\bm{p}}
  = I_{\parallel} + \bm{q} \bm{q} / \gamma_0^2 ,
\end{equation}
where $I_{\parallel}$ is the identity matrix in the transverse plane with
\begin{equation*}
  I_{\parallel} = 
  \begin{bmatrix}
    1 & 0 & 0 \\
    0 & 1 & 0 \\
    0 & 0 & 0
  \end{bmatrix} .
\end{equation*}
Thus, the effective self-energy takes the final form
\begin{equation} \label{Pi_infty_2D}
  \Pi_{\infty}^r(\bm{q}) = \big[ \Pi_{\infty}^a(\bm{q})\big]^* = 
  -i \epsilon_0 c^2 \gamma_0 \big( I_\parallel + \bm{q} \bm{q} / \gamma_0^2 \big) ,
\end{equation}
which constitutes the second central result of this work.
Using Eq.~\eqref{Xi_2D_surface}, the environmental contribution in Eq.~\eqref{X_en}
simplifies to
\begin{equation} \label{X_en_infty_2D}
  X_{\rm en}(z, z') = D^r(z, z_\infty) \big( \Pi_\infty^r - \Pi_\infty^a \big)
  D^a(z_\infty, z'),
\end{equation}
where the bare Green's functions have been replaced by the full ones to account for the
presence of the object.

Several remarks are in order regarding Eq.~\eqref{Pi_infty_2D}. 
First, as in the three-dimensional case, we must attach the transverse projector
$P_\parallel$, since the projectors already present in $D_0^r(z, z_\infty)$ and
$D_0^a(z_\infty, z')$ can be absorbed or canceled by the object's self-energy when
simplifying Eq.~\eqref{X_en} to Eq.~\eqref{X_en_infty_2D}.
Second, while the three-dimensional bath self-energy in Eq.~\eqref{Pi_infty_3D} scales
linearly with frequency and is Ohmic, the planar projection introduces a geometric
frequency dependence via $\gamma_0$.
Third, this effective self-energy is independent of the chosen plane $z_{\infty}$, because
any spatial shift $\Delta z$ introduces a phase factor $e^{i\gamma_0 \Delta z}$ to the
retarded Green's function $D_0^r$ and its complex conjugate $e^{-i\gamma_0 \Delta z}$ to
$D_0^a$, which exactly cancel in Eq.~\eqref{Xi_2D_surface}. 
Fourth, for an object of finite thickness, one needs to include photonic baths at both
positive and negative infinities to account for radiation emitted into both half-spaces. 
Finally, although Eq.~\eqref{Pi_infty_2D} diverges at the light line ($q \to k_0$,
$\gamma_0 \to 0$), this singularity is integrable. Any macroscopic physical observable
$I(\omega)$ requires integration over the wavevector space, $d^2\bm{q} = q\, dq\, d\phi$,
taking the general form with
\begin{equation*}
  I(\omega) \propto \int_{0}^{k_0} F(q, \omega) \frac{1}{\gamma_0} \, q dq 
  = \int_{0}^{k_0} q dq \frac{ F(q, \omega)}{\sqrt{k_0^2 - q^2}} ,
\end{equation*}
where $F(q, \omega)$ encapsulates the system's well-behaved physical response. Through the
variable change $x = k_0^2 - q^2$ so that $dx = -2q\,dq$, the singularity is integrated
out, yielding a term proportional to $\sqrt{x}$. Therefore, this integral evaluates to a
finite value as long as $F(k_0, \omega)$ is finite.

\subsection{Bath at infinity behaves as a black body}
The expression of the energy density at $z=0$ in Eq.~\eqref{u_D} can be expressed as
\begin{equation}
  u = 2 \int \frac{d^2\bm{q}}{(2\pi)^2} \int_0^\infty \frac{d\omega}{\pi} i\hbar
  \epsilon_0 \omega^2 \, {\rm tr} \big[ D_0^<(\bm{q}, z, z) \big]_{z=0} ,
\end{equation}
where we have used the fact that the electric and magnetic contributions are equal, and
the factor $2$ accounts for the photonic baths at both positive and negative infinities.
With in-plane translational invariance, the lesser component of the bare Green's function
satisfies
\begin{equation}
  D_0^<(\bm{q}, z, z') = D_0^r(\bm{q}, z, z_\infty) \Pi_\infty^<(\bm{q})
  D_0^a(\bm{q}, z_\infty, z'),
\end{equation}
where $\Pi_\infty^<$ is given by Eq.~\eqref{Pi<_infty_FD}.
Using the far-field approximations in Eq.~\eqref{D0f_2D}, the form of
$\Pi_{\infty}^r$ in Eq.~\eqref{Pi_infty_2D}, the relation in Eq.~\eqref{PPP}, and the
trace identity ${\rm tr}(\mathbb{P} ) = 2$, we obtain 
\begin{equation}
  u = 2\int_0^\infty \frac{d\omega}{\pi} \int \frac{d^2\bm{q}}{(2\pi)^2} \frac{\hbar
  \omega^2}{c^2 \gamma_0} N_0(\omega) .
\end{equation}
Performing the angular integration over $\bm{q}$, we have
\begin{equation}
  \int \frac{d^2\bm{q}}{(2\pi)^2} \frac{1}{\gamma_0} = \frac{1}{2\pi} \int_0^{k_0}
  \frac{qdq}{\gamma_0} = \frac{k_0}{2\pi} .
\end{equation}
Substituting this into the expression for $u$ gives the blackbody radiation energy density
in Eq.~\eqref{Planck}.

\subsection{Far-field thermal radiation} 
We discuss the implication of photonic bath at infinity on the far-field thermal radiation
from a semi-infinite object with in-plane translational invariance.
Given that the total heat current $J$ is proportional to the interface area, we define the
radiative heat flux as $j = J / S_\parallel$ with $S_\parallel$ the infinite transverse
area. Assuming thermal equilibrium for the object, the heat flux is expressed as
\begin{align} 
  j = -2i \int_0^{\infty} \frac{d\omega}{2\pi} & 
  \int_{q<k_0} \frac{d^2\bm{q}}{(2\pi)^2} \hbar\omega \int_{-\infty}^0 dz {\rm tr} 
  \big\{ {\rm Im} \big[\Pi_1^r(\bm{q}, z) \big]  \notag \\
  & X_{\rm en}(\bm{q}, z, z)\big\} [N_1(\omega) - N_0(\omega)] , 
\end{align}
where the optical response is assumed to be local in the $z$-direction with 
$\Pi_1^r(\bm{q}, z',z) = \Pi_1^r(\bm{q}, z) \delta(z-z')$, 
and $X_{\rm en}$ is given by Eq.~\eqref{X_en_infty_2D}.
The integration over the in-plane wavevector $\bm{q}$ is restricted to the far-field
region with $q < k_0$.
This expression can be further simplified into a Landauer-B\"{u}ttiker form as
\begin{equation}
  j = \int_0^{\infty} \frac{d\omega}{2\pi} \hbar\omega \int_{q<k_0}
  \frac{d^2\bm{q}}{(2\pi)^2} \xi(\bm{q}, \omega) [N_1(\omega) - N_0(\omega)] ,
\end{equation}
where the photonic transmission function is
\begin{equation} \label{xi}
  \xi = 4 \int_{-\infty}^0 dz  {\rm tr} \big\{ D^a(z_\infty, z) {\rm Im}[\Pi_1^r(z)]
  D^r(z, z_\infty) {\rm Im}(\Pi^r_\infty) \big\} ,
\end{equation}
with the variables of $\bm{q}$ and $\omega$ being omitted. 

Using Eq.~\eqref{optical_22}, the photonic transmission function can be rewritten in terms
of the Green's functions evaluated at $(z_\infty, z_\infty)$ as
\begin{equation} \label{xi2}
  \xi = -2i {\rm tr} \big\{ \big[ D^r - D^a - 2i D^a {\rm Im}(\Pi^r_\infty)
  D^r \big] 
{\rm Im}(\Pi^r_\infty)  \big\} .
\end{equation}
Following the method of Ref.~\cite{Sipe87}, the retarded Green's function at the far-field
location $z_\infty$ is
\begin{equation} 
  D^r(z_\infty , z_\infty) = v(0) + \Upsilon
  \begin{bmatrix}
    \hat{\bm{s}}  &  \hat{\bm{p}}_{+}
  \end{bmatrix}
  \mathsf{R}_1
  \begin{bmatrix}
    \hat{\bm{s}}^T  \\
    \hat{\bm{p}}_{-}^T
  \end{bmatrix}
  e^{2i\gamma_0 z_\infty} ,
\end{equation}
where $\mathsf{R}_1$ is the reflection matrix for waves incident from the vacuum onto the
interface. Since $\Pi^r_\infty$ acts as a two-by-two matrix in the transverse plane, we
use the in-plane projections 
$\mathsf{D}^r = \big[ D^r(z_\infty,z_\infty) \big]_{\parallel}$ and 
$\mathsf{D}^a = \big[ \mathsf{D}^r \big]^\dag$ in Eq.~\eqref{xi2}.
They are expressed as 
\begin{align}
  & \mathsf{D}^r = \Upsilon \mathsf{V} \big(\mathsf{I} + \sigma_z \mathsf{R}_1
  e^{2i\gamma_0 z_\infty} \big) \mathsf{V}^T ,  \\
  & \mathsf{D}^a = -\Upsilon \mathsf{V} \big(\mathsf{I} + \mathsf{R}_1^\dag \sigma_z
  e^{-2i\gamma_0 z_\infty} \big) \mathsf{V}^T ,
\end{align}
with
\begin{equation}
  \mathsf{I} = 
  \begin{bmatrix}
    1 & 0 \\ 0 & 1
  \end{bmatrix} , \qquad
  \sigma_z = 
  \begin{bmatrix}
    1 & 0 \\ 0 & -1
  \end{bmatrix} , \qquad
  \mathsf{V} = 
  \begin{bmatrix}
    \hat{\bm{s}} & \hat{\bm{p}}
  \end{bmatrix}, 
\end{equation}
where the reduced polarization vectors are
\begin{equation}
  \hat{\bm{s}} = \frac{1}{q} 
  \begin{bmatrix}
    q_y \\  -q_x
  \end{bmatrix} , \qquad
  \hat{\bm{p}} = \frac{\gamma_0}{qk_0} 
  \begin{bmatrix}
    q_x \\  q_y
  \end{bmatrix} .
\end{equation}
The reduced polarization vectors satisfy the relations:
\begin{equation} \label{VF}
  \mathsf{V} \mathsf{F} \mathsf{V}^T = \mathsf{V}^T \mathsf{F} \mathsf{V} = \mathsf{I} ,
  \qquad
  \mathsf{V} \mathsf{V}^T = \mathsf{V}^T \mathsf{V} = \mathsf{F}^{-1} ,
\end{equation}
with
\begin{equation}
  \mathsf{F} = 
  \begin{bmatrix}
    1 & 0 \\ 0 & k_0^2 / \gamma_0^2 
  \end{bmatrix} .
\end{equation}
Given the self-energy of the bath at infinity 
\begin{equation}
  \Pi_\infty^r = - \frac{1}{2\Upsilon} \mathsf{V}\mathsf{F}^2 \mathsf{V}^T ,
\end{equation}
a straightforward calculation of Eq.~\eqref{xi2} yields
\begin{equation}
  \xi(\bm{q}, \omega) = {\rm tr} \big(\mathsf{I} - \mathsf{R}_1^\dag \mathsf{R}_1 \big) .
\end{equation}
Consequently, the heat flux is simplified to~\cite{Kruger12}
\begin{equation}
  j = \int_0^{\infty} \frac{d\omega}{2\pi} \hbar\omega \int_{q<k_0}
  \frac{d^2\bm{q}}{(2\pi)^2} {\rm tr} \big(\mathsf{I} - \mathsf{R}_1^\dag \mathsf{R}_1
  \big) [N_1(\omega) - N_0(\omega)] ,
\end{equation}
where $\mathsf{R}_1$ represents the generalized Fresnel reflection matrix, accounting for
potential cross-polarization conversion.

\section{Alternative derivation via surface Green's function approach} \label{SecV}
In this section, we provide an alternative derivation of the bath self-energy using the
surface Green's function technique introduced in Ref.~\cite{Jiebin17}.
Notably, this approach is independent of the dust model proposed in Ref.~\cite{Eckhardt84}
and offers a distinct theoretical perspective.

We first consider the planar geometry. Due to in-plane rotational symmetry, we set $q_x =
0$ and $q_y = q$ without loss of generality. In this representation, Eq.~\eqref{A_EOM_eta}
reduces to
\begin{equation}
  \epsilon_0 c^2 \big[(\tilde{\gamma}_0^2 + \partial_z^2)\mathbb{I} + \bm{Q} \bm{Q}\big]
  D_0^r(z,z') = \mathbb{I} \delta(z-z') , 
\end{equation}
where $\tilde{\gamma}_0$ is given in Eq.~\eqref{tilde_gamma_0}, and 
$\bm{Q} = (0, q, -i\partial_z)$.
This dyadic equation decouples into independent components for the different
polarizations.
Specifically, the $xx$ and $yy$ components satisfy
\begin{align}
  &\epsilon_0 c^2 (\tilde{\gamma}_0^2 + \partial_z^2) D_{0, xx}^r(z,z') = \delta(z-z') , 
  \label{D0xx} \\
  &\epsilon_0 c^2 (\tilde{\gamma}_0^2 + \partial_z^2) D_{0, yy}^r(z,z') = 
  (\gamma_0^2 /k_0^2) \delta(z-z').
\label{D0yy}
\end{align}
With the choice of $q_x=0$, a comparison between the dyadic components and the
polarization vectors in Eq.~\eqref{sp} confirms that $D_{0, xx}^r$ and $D_{0, yy}^r$
correspond to the $s$- and $p$-polarized modes, respectively.

\begin{figure}
\centering
\includegraphics[width=2.6in]{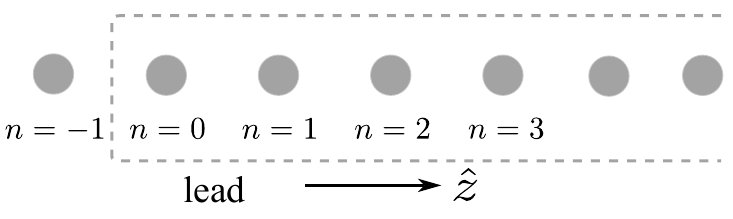} \\
\caption{Schematic of the discrete lattice used for the surface Green's function
  derivation. The indices $n \ge 0$ denote the semi-infinite vacuum lead, which
  collectively acts as the effective photonic bath at infinity coupled to the system
  interface at $n = -1$. }
\label{fig2}
\end{figure}

To implement the surface Green’s function approach, we discretize the $z$-axis into a
lattice with grid spacing $a$, where $z_n = n a$ (see Fig.~\ref{fig2}).
Using the central difference approximation for the second derivative, Eq.~\eqref{D0xx} at
grid point $n$ becomes
\begin{equation}
  \epsilon_0 c^2 \bigg( \frac{D_{n+1} - 2D_n + D_{n-1}}{a^2} + \tilde{\gamma}_0^2 D_n
  \bigg) = \frac{\delta_{n0}}{a} ,
\end{equation}
where the source $z'$ is located at $n=0$.
We define the hopping parameter $V$ and the on-site ``energy" $\varepsilon$ as
\begin{equation} \label{V_E}
  V = \epsilon_0 c^2 / a^2 ,\quad \varepsilon = V (\tilde{\gamma}_0^2 a^2 - 2) .
\end{equation}
The discrete equation then takes the form of a standard tight-binding model:
\begin{equation} \label{Dn}
  V D_{n-1} + \varepsilon D_n + V D_{n+1} = \delta_{n0} /a  .
\end{equation}
Following the method in Refs.~\cite{Datta95, JSW07, JSW14}, we treat the vacuum for $n \ge
0$ as a semi-infinite lead coupled to a central system (see Fig.~\ref{fig2}). For $n \ge
0$, the homogeneous lead equations admit the ansatz 
\begin{equation}
  D_n = \Lambda \lambda^n .
\end{equation}
In the limit $a \to 0$, the characteristic equation yields
\begin{equation}
  \lambda = \frac{-\varepsilon \pm \sqrt{\varepsilon^2 -4V^2}}{2V} 
  = 1 \pm i\tilde{\gamma}_0 a - \frac{1}{2} \tilde{\gamma}_0^2 a^2 
  = e^{\pm i\tilde{\gamma}_0 a} .
\end{equation}
To satisfy the boundary condition for $n \to \infty$, we choose
\begin{equation}
  \lambda = \exp{[i\gamma_0 a - \eta k_0 a /(c\gamma_0)]} .
\end{equation}
Applying the boundary condition $\varepsilon D_0 + V D_1 = 1 / a$, we find 
$D_0= \Lambda = 1/ [a(\varepsilon + V\lambda ) ]$.
The self-energy $\Sigma_s^r$, which accounts for the influence of the semi-infinite lead
on the system interface ($n=-1$), is given by 
\begin{equation} \label{Sigma_s_0}
  \Sigma_s^r = a^2 V D_0 V = a V^2 / (\varepsilon + V\lambda) .
\end{equation}
In the continuum limit $a \to 0$, we obtain 
\begin{equation} 
  \Sigma_s^r = - \epsilon_0 c^2 / a - i \epsilon_0 c^2 \gamma_0 .
\end{equation} 
The real part of this expression exhibits an unphysical static divergence representing an
infinite shift in the system's potential energy.
Physically, the inverse lattice spacing $1/a$ acts as a high-momentum cutoff,
corresponding to the high-frequency cutoff $\omega_D$ in the Drude model for Ohmic
baths~\cite{Weiss}.
To eliminate this potential renormalization, the Caldeira-Leggett
framework~\cite{Caldeira-81, Weiss} introduces a counter-term.
This ensures that the bath solely introduces dissipation without altering the system's
potential or breaking translational invariance. Consequently, the properly renormalized
self-energy retains only the physically meaningful dissipation with
\begin{equation} \label{Sigma_s}
  \Sigma_{s,{\rm ren}}^r = - i \epsilon_0 c^2 \gamma_0 .
\end{equation}
Furthermore, the continuum limit $a \to 0$ implies the Markovian nature of the bath at
infinity, since a finite lattice spacing implies a finite bandwidth and finite-time memory
effects.

For $p$-polarized mode, the discrete form of Eq.~\eqref{D0yy} includes the factor
$\gamma_0^2/k_0^2$ on the right-hand side. This indicates that the hopping parameter and
on-site ``energy" in Eq.~\eqref{V_E} are modified by the factor $k_0^2 / \gamma_0^2$.
Consequently, a comparison with Eq.~\eqref{Sigma_s_0} for the $s$-polarization mode shows
that the effective $p$-mode self-energy scales as
\begin{equation}
  \Sigma_{p,{\rm ren}}^r = \frac{k_0^2}{\gamma_0^2} \Sigma_{s,{\rm ren}}^r .
\end{equation}
Combining these results, the dyadic self-energy in the planar system is
\begin{equation}
  \Pi_{\infty}^r = \Sigma_{s,{\rm ren}}^r \hat{\bm{x}} \hat{\bm{x}} + 
                   \Sigma_{p,{\rm ren}}^r \hat{\bm{y}} \hat{\bm{y}} .
\end{equation}
Using the planar rotational symmetry, the self-energy is then
\begin{equation}
  \Pi_{\infty}^r(\bm{q}) = -i \epsilon_0 c^2 \gamma_0 
\big[\hat{\bm{s}} \hat{\bm{s}} + (k_0 / \gamma_0)^4 \hat{\bm{p}} \hat{\bm{p}} \big],
\end{equation}
which is in exact agreement with the effective self-energy derived via the dust model in
Eq.~\eqref{Pi_2D_infty}.


For the three-dimensional case, we set $q=0$ such that $\gamma_0 = k_0$.
The self-energy
in Eq.~\eqref{Sigma_s} reduces to
\begin{equation}
  \Sigma_{3D}^r = - i \epsilon_0 c \omega .
\end{equation}
By applying the projection operator $\mathbb{I} - \hat{\bm{R}} \hat{\bm{R}}$, we recover
the self-energy given in Eq.~\eqref{Pi_infty_3D}, confirming the universality of the
approach based on the lattice Green's function.

\section{Summary} \label{SecVI}
The theoretical formalism of the photonic bath at infinity has been systematically
developed using the nonequilibrium Green's function technique.
Despite the infinitesimal nature of the environmental ``dust", integration over infinite
space yields a finite result, which can be represented by an effective self-energy
describing the photonic bath at infinity. 
Our approach eliminates the need for cumbersome volume meshing of the infinite space in
numerical computations by replacing the environment with a simple boundary condition.
Analytical expressions for this bath self-energy have been derived for both finite
three-dimensional objects and planar systems with in-plane translational invariance, with
the main results given in Eqs.~\eqref{Pi_infty_3D} and \eqref{Pi_infty_2D}. 
The bath is further demonstrated to behave as a black body, and its role in far-field
thermal radiation is elucidated. The formalism is readily extendable to time-dependent
driving and more complex geometries. For periodically time-modulated system, because the
vacuum is unmodulated, the bath self-energy is diagonal in Floquet space with
$\big[\Pi_\infty^r\big]_{mn}(\omega) = \Pi_\infty^r(\omega_m)\delta_{mn}$. Furthermore,
the planar bath framework applies to transverse periodic structures, such as metasurfaces,
provided they possess in-plane translational invariance.

\begin{acknowledgments}
G.T. acknowledges support from Science Challenge Project (Grant No. TZ2025017) and
National Natural Science Foundation of China (Grants No. 12088101 and No. 12374048).
\end{acknowledgments}

\appendix
\numberwithin{equation}{section}

\section{Symmetries of the photonic Green's functions} \label{Appendix_A}
\renewcommand{\theequation}{A.\arabic{equation}}
From the definition of the Green's function, we have
\begin{equation} 
  \big[ D^r(j,i) \big]^\dag = D^a(i,j) , \quad
  \big[ D^{<,>}(j,i) \big]^\dag = -D^{<,>}(i,j) .
\end{equation}
where the indices $i$ and $j$ denote both spatial coordinates and time arguments.
In the frequency domain, these yield 
\begin{align} 
  & \big[D^r(\bm{r}_1, \bm{r}_2, \omega)\big]^\dag =  D^a(\bm{r}_2, \bm{r}_1, \omega),
  \label{symD_11} \\
  & \big[D^<(\bm{r}_1, \bm{r}_2, \omega)\big]^\dag = -D^<(\bm{r}_2, \bm{r}_1, \omega).
\label{symD_12}
\end{align}
Additionally, using the bosonic commutation relation for the vector potential, we obtain
further symmetries with
\begin{equation} 
  \big[ D^r(j,i) \big]^T = D^a(i,j) , \quad
  \big[ D^>(j,i) \big]^T = D^<(i,j) .
\end{equation}
In the frequency domain, these relations become
\begin{align} 
  D_{\mu\nu}^r(\bm{r}_1, \bm{r}_2, \omega) &= D_{\nu\mu}^a(\bm{r}_2, \bm{r}_1,-\omega) ,
  \label{symD_21} \\
  D_{\mu\nu}^>(\bm{r}_1, \bm{r}_2, \omega) &= D_{\nu\mu}^<(\bm{r}_2, \bm{r}_1,-\omega) .
\label{symD_22}
\end{align}
For systems with in-plane translational invariance, the symmetries in
Eqs.~\eqref{symD_11}, \eqref{symD_12}, \eqref{symD_21}, and \eqref{symD_22} become
\begin{align}
  & \big[ D^r(\bm{q}, \omega, z, z') \big]^\dag = D^a(\bm{q}, \omega, z', z),
   \label{symD_q11} \\
  & \big[ D^<(\bm{q}, \omega, z, z') \big]^\dag = -D^<(\bm{q}, \omega, z', z),
   \label{symD_q12} \\
  & D_{\mu\nu}^r(\bm{q}, \omega, z, z') = D_{\nu\mu}^a(-\bm{q}, -\omega, z', z),
   \label{symD_q21} \\
  & D_{\mu\nu}^>(\bm{q}, \omega, z, z') = D_{\nu\mu}^<(-\bm{q}, -\omega, z', z).
\label{symD_q22}
\end{align}
All the above symmetries also hold for the self-energy $\Pi$.

\bibliography{bib_heat_radiation}

\end{document}